# Multiple Superconducting Phases in Rhombohedral Heptalayer Graphene

Chuanqi Zheng[1†], Chushan Li[1†], Chenyu Zhang[1], Kenji Watanabe[2], Takashi Taniguchi[3], Hao Yang [1,4,5], Dandan Guan[1,4,5], Liang Liu[1,4,5], Shiyong Wang[1,4,5], Yaoyi Li[1,4,5], Hao Zheng[1,4,5], Canhua Liu [1,4,5], Jinfeng Jia[1,4,5,6,7], Zhiwen Shi[1], Guorui Chen[1], Tingxin Li[1,4*] and Xiaoxue Liu[1,4,5*]

[1]State Key Laboratory of Micro-nano Engineering Science, Tsung-Dao Lee Institute & School of Physics and Astronomy, Key Laboratory of Artificial Structures and Quantum Control (Ministry of Education), Shanghai Jiao Tong University, Shanghai 200240, China

[2]Research Center for Electronic and Optical Materials, National Institute for Materials Science, 1-1Namiki, Tsukuba, Japan

[3]Research Center for Materials Nanoarchitectonics, National Institute for Materials Science, 1-1 Namiki, Tsukuba, Japan

[4]Hefei National Laboratory, Hefei 230088, China

[5]Shanghai Research Center for Quantum Sciences, 99 Xiupu Road, Shanghai 201315, China

[6]Quantum Science Center of Guangdong-Hong Kong-Macao Greater Bay Area (Guangdong), Shenzhen 518045, China

[7]Department of Physics, Southern University of Science and Technology, Shenzhen 518055, China

[†]These authors contributed equally to this work.

[*]Emails: txli89@sjtu.edu.cn, xxliu90@sjtu.edu.cn

**Key words:** Rhombohedral multilayer graphene; Superconductivity; Quantum transport

**Abstract**

**Crystalline rhombohedral multilayer graphene (RMG) has emerged as an ideal platform for studying unconventional superconductivity. Here, we report the observation of superconductivity in moiréless rhombohedral heptalayer graphene (RHG) at zero magnetic field. The superconducting phases emerge at low displacement electric fields ($|D| < 0.2$ V/nm) and are symmetrically distributed about $D = 0$, with one robust state exhibiting zero resistance and two weaker superconducting features. Comparisons with rhombohedral pentalayer graphene (RPG) reveal distinct perpendicular magnetic-field responses, and quantum oscillation measurements indicate that superconductivity in RHG arises from a half-metallic normal state. These results highlight the strong dependence of superconductivity on layer number and electronic structure in RMG systems and provide new insights into its microscopic origin.**

Since the discovery of superconductivity in magic-angle twisted bilayer graphene [1], superconductivity has been observed in a variety of graphene-based systems [2–34]. Recently, superconductivity in moiréless Bernal bilayer graphene and RMG has attracted considerable attention due to its exotic properties, including time-reversal symmetry breaking [24,26], pronounced violations of the paramagnetic Pauli limit [16-24,29,30,31,33,34], and superconductivity induced by an in-plane magnetic field [17,30-34]. RMG has a simple chemical composition, consisting entirely of carbon atoms. In the simplest case, the low-energy band dispersion in RMG follows the relation of $E \propto k^N$ (where $N$ is the number of layers), resulting in a significantly enhanced density of states (DOS) near the band edge, which strongly amplifies electron-electron interactions. Moreover, by imposing an interlayer chemical potential difference, the band structure of RMG can be effectively tuned through electrostatic gating. Together, these features make RMG an ideal and highly tunable platform for experimentally investigating unconventional superconductivity and other emergent phenomena arising from electron correlations. So far, robust superconductivity has already been reported in moiréless rhombohedral-stacked trilayer [16,22,23,31], tetralayer [24,30], pentalayer [24,30], hexalayer [26,27,29,34], octalayer [28], and thirteen-layer graphene [33], with the notable exception of heptalayers.

In this work, we report the first observation of superconductivity in moiréless RHG at zero magnetic field, thereby completing the superconducting sequence for layer numbers below eight. These states appear within a displacement electric field range of $0 < |D| < 0.2$ V/nm and are symmetrically distributed about $D = 0$, with three states at positive electric fields and three at negative electric fields. By comparing the results with those of RPG [30], we find that although both systems host rich superconducting phase diagrams, their characteristics are distinct. Notably, the perpendicular-magnetic-field-enhanced superconductivity observed in RPG is absent in RHG. Furthermore, Fermi surface analysis indicates that the normal state of the superconducting phase in RHG corresponds to a half metallic phase with twofold degeneracy.

Figures 1(a) and 1(b) show the schematic and the optical image of the device, respectively. The device consists of RHG equipped with dual-graphite gates, enabling independent control of the carrier density $n$ and the perpendicular displacement electric field $D$. These parameters are determined by the relations $n = (C_{top}V_{top} + C_{bottom} V_{bottom})/e + n_0$ and $D = (C_{top}V_{top} - C_{bottom} V_{bottom})/(2\varepsilon_0) + D_0$, where $C_{top(bottom)}$ and $V_{top(bottom)}$ are the geometric capacitances and applied voltages for the top (bottom) gates, respectively, $n_0$ represents the intrinsic doping, $\varepsilon_0$ is the vacuum permittivity, and $D_0$ denotes the built-in displacement electric field. Transport measurements were performed in a dilution refrigerator (Q-one, model Q-400) equipped with a 9 T superconducting magnet, using standard low-frequency lock-in techniques with an excitation current of 1-3nA at a frequency of 17.7 Hz. Each fridge line has a silver epoxy filter and multiple stage RC - filters at low temperature.

Figure 1(c) shows the $n$-$D$ phase diagram of the longitudinal resistivity $\rho_{xx}$ at zero magnetic field. At the charge neutrality point, the system transitions from a layer antiferromagnetic insulator at low electric fields to a semimetallic state, and eventually to a layer-polarized insulating state at higher displacement electric fields, consistent with previous reports on RMG [24-39]. Notably, three symmetric pairs of superconducting-like features are observed about $D = 0$ on the hole-doped side, labeled as SC1, SC2' and SC3' in Fig. 1(d). Owing to the symmetry across $D = 0$, the same labels are used to denote the equivalent superconducting-like states in both the $D > 0$ and $D < 0$ regimes. Among these, SC1 occupies a larger area in the phase diagram, showing a curved feature near the boundary of the multiferroicity regime [26,29,30,34,35,38]. The other two states, SC2' and SC3', occupy smaller regions flanking SC1.

To further characterize the properties of these states, Figs. 2(a) - (c) show the temperature dependence of $\rho_{xx}$ as a function of carrier density $n$ near the three superconducting states, respectively. At the lowest temperature of our refrigerator (nominally 10 mK), only SC1 is fully developed and reaches zero resistivity within the noise floor. In contrast, SC2' and SC3' exhibit finite residual resistivity, though both show a strong temperature dependence with resistivity decreasing sharply upon cooling. The highest critical temperature $T_c$ for SC1 in Fig. 2(a) is about 134 mK, defined as the temperature where the resistivity reaches 50% of its normal-state resistivity. Figures 2(d) - 2(f) show the perpendicular magnetic field dependence of the differential resistivity d$V$/d$I$ as a function of the dc current $I_{dc}$ for SC1, SC2' and SC3', respectively. The differential resistivity d$V$/d$I$ of SC1 exhibits pronounced nonlinear behavior and signatures of Fraunhofer-like interference pattern, providing a hallmark signature of the superconducting phase. In comparison, SC2' and SC3' also display nonlinear d$V$/dI characteristics, although they are less pronounced than those observed for SC1. Figures S2(a) – S2(c) show the map of $\rho_{xx}$ as a function of density and perpendicular magnetic field $B_\perp$ for SC1, SC2' and SC3', respectively. Consistent with the critical temperature measurements, SC1 is the most robust and exhibits the highest critical perpendicular magnetic field of about $B_{c\perp} = 12.5$ mT, whereas SC2' and SC3' are relatively weak and are fully suppressed by $B_\perp$ below 1 mT. All these states are strongest at zero perpendicular magnetic field and are gradually suppressed with increasing $B_\perp$.

Combining these experimental observations of SC2' and SC3' states that the resistivity drops upon cooling, the robust nonlinear differential resistivity, and the complete suppression of these states by applying a relatively small perpendicular magnetic field on the order of mT, we conclude that the SC2' and SC3' states are consistent with superconductivity. Strictly speaking, both states exhibit a finite residual resistivity at the base temperature (about 10 mK). Similar behaviors have been widely reported in superconducting states of RMG systems [22-23,25,28-30,32-34], particularly for those with low critical temperatures (on the order of tens of millikelvin). These fragile superconducting states are more susceptible to disorder or subtle heating effects, resulting in non-zero resistance even at the base temperature. Taken together, we assign SC2' and SC3' states as signatures of superconductivity, although further measurements would be needed to establish their superconducting nature more rigorously.

It is important to note that although the superconductivity can be observed in rhombohedral graphene with different layer numbers, its properties show a strong dependence on both the number of layers and the displacement electric field [16,22-24,26-30,31,33,34]. We next compare the superconducting properties emerging at low displacement electric fields in RHG with those measured in a RPG (Fig. S3). The measurements were performed under comparable conditions for both systems, enabling a direct comparison of their superconducting states (see

Table I in the Supplementary Material). The *n*-*D* phase diagram of $\rho_{xx}$ (Figs. 3(a) - 3(b)) in RPG reveals six superconducting-like states distributed symmetrically about $D = 0$, consistent with observations in earlier RPG studies [30]. We focus on the superconductivity in the negative *D*-field regime, labeling them as $SC1_5$, $SC2_5$', and $SC3_5$, respectively. The positions of $SC1_5$, $SC2_5$', and $SC3_5$ in the *n*-*D* phase diagram are different from those in RHG. This distinction may be linked to band structures or to distinct isospin-symmetry-breaking configurations in RMG systems with different layer numbers. The microscopic origin is likely related to the underlying pairing mechanisms, which is beyond the scope of the present work and warrants further theoretical and experimental investigations.

The temperature dependence of resistivity for the $SC1_5$, $SC2_5$'and $SC3_5$ states is shown in Fig. S4, and the critical temperatures of $SC1_5$ and $SC3_5$ are on the same order of magnitude as those in RHG. Considering the finite residual resistivity of $SC2_5$' at the base temperature, we refer to the $SC2_5$' state as a signature of superconductivity. Figure 3(c) - 3(e) shows the map of differential resistivity $dV/dI$ as a function of direct current $I_{dc}$ and perpendicular magnetic field $B_\perp$. Remarkably, similar to previous observations [30], the evolution of $SC3_5$ exhibits an unusual non-monotonic trend as a function of $B_\perp$. Specifically, the superconducting state of $SC3_5$ is enhanced with increasing perpendicular magnetic fields, reaching its maximum at a small finite magnetic field of approximately 1.6 mT, before being suppressed as the magnetic field is further increased. On the other hand, $SC1_5$ and $SC2_5$' become suppressed monotonically by increasing the external perpendicular magnetic field, similar to the behavior of SC1, SC2' and SC3' observed in RHG. As shown in Figs. S2(d) - (f), the $B_\perp$ dependence of $\rho_{xx}$ versus carrier density *n* is consistent with that of $dV/dI$ measurements. The superconducting density ranges for $SC1_5$ and $SC2_5$' are broadest at zero magnetic field, whereas $SC3_5$ reaches its maximum density range only under a finite perpendicular magnetic field $B_\perp$. The $B_\perp$ field-induced enhancement of the $SC3_5$ state is also evident by comparing the *n*-*D* phase diagrams at $B_\perp = 0$ and $B_\perp = 1.6$ mT (Figs. 3(f) and 3(g)). Obviously, the $SC3_5$ state expands significantly upon the application of the $B_\perp$ field. This enhancement is further manifested by critical temperature measurements shown in Fig. 3(h)- 3(i). Specifically, at the optimal doping for $D = -0.061$ V/nm, $T_c$ increases from 83 mK ($B_\perp = 0$) to 100 mK ($B_\perp = 1.6$ mT; Fig. 3(j)), underscoring the strengthening of the $SC3_5$ phase by a finite $B_\perp$ field.

To provide further insights into the pairing mechanisms of superconductivity, we analyze the fermiology of the parent normal state. In particular, we perform quantum oscillation measurements to investigate the possible isospin configurations or Fermi pocket structures of the normal states. Figure 4(a) displays $\rho_{xx}$ as a function of $B_\perp$ and *n* at a fixed $D = 0.157$ V/nm, in the vicinity of the SC1 state in RHG. The observed quantum oscillations are clearly resolved through Fast Fourier transform (FFT) analysis of $\rho_{xx}$ ($1/B_\perp$), as shown in Fig. 4(b). Here, the normalized frequency $f_v$ represents the fraction of the total Fermi surface area enclosed by a cyclotron orbit and is given by $f_v = f_{FFT} \times e/nh$, where $f_{FFT}$ is the quantum oscillation frequency of $\rho_{xx}$ ($1/B_\perp$) extracted from the FFT spectra, *e* is the electron charge, and *h* is Planck's constant. According to Luttinger's theorem, the sum of $f_v^{(i)}$ weighted by the degeneracy $D_i$ should be normalized, i.e., $\sum_i(\pm D_i f_v^{(i)}) = 1$, where the sign is positive (+) or negative (-) depending on whether the doping charge carriers have the same or opposite charge polarity.

As shown in Figs. 4(a) and 4(b), the parent normal state of SC1 is consistent with a half-metallic phase with two-fold degeneracy. This is supported by two frequency peaks satisfying the relation $|f_v^{(1)} - f_v^{(2)}| = 1/2$, which could arise from either a two-fold degenerate annular Fermi surface or a semimetallic phase consisting of two electron pockets and two hole pockets[38]. Nevertheless, the precise fermiology of the normal state cannot be uniquely resolved from quantum oscillations alone and remains to be determined by further experimental studies. Upon decreasing the charge carrier density, the dominant frequencies transition to satisfy $f_v^{(1)} + f_v^{(2)} = 1$, suggesting an evolution into a partially isospin-polarized phase, likely consisting of one major and one minor Fermi pocket (denoted as $PIP_1$). Consequently, the observed SC1 phase in RHG emerges from a half-metallic normal state near the boundary of the $PIP_1$ phase. In contrast, the normal state fermiology of RPG is illustrated in Fig. S5. The normal state of SC1 likely corresponds to an annular Fermi surface with fourfold degeneracy. As for $SC2_5$' and $SC3_5$ states, which emerge at positions in the phase diagram similar to the SC1 observed in RHG, their normal states similarly correspond to a half-metallic phase, however, no partially isospin-polarized phase is observed in this case. Taken together, although the pairing mechanism remains elusive [40-55], the Fermi surface structures of the normal state provide essential clues regarding the origin of superconductivity in RMG.

In summary, we observed the superconducting states in moiréless RHG at zero magnetic field, symmetrically distributed with respect to $D = 0$. Compared with the superconductivity observed in RHG/hBN moiré superlattices [28,32], the superconducting states in moiréless RHG emerge at different locations in the $n$-$D$ phase space, suggesting that the moiré potential plays a role in shaping superconductivity in RMG/hBN systems rather than being directly inherited from moiréless RHG.

Notably, the superconducting states in moiréless RHG appear in a similar region of the $n$-$D$ phase diagram to those previously reported in moiréless RMG systems with layer numbers ranging from five to eight [28, 29, 30, 34, 38]. This region corresponds to a low displacement electric field regime with $|D| < 0.2$ V/nm, and has been identified as a dual-surface semimetallic phase, in which electron and hole carriers coexist and are spatially separated, residing predominantly in the topmost and bottommost layers, respectively. Although the specific properties of these superconducting states can differ, their emergence in a similar region of the $n$–$D$ phase diagram suggests that these superconducting states may share a common or related origin. In contrast, the $SC1_5$ phase emerges out of the semimetallic region. A similar superconducting state has also been observed in rhombohedral tetralayer graphene at analogous positions in the $n$-$D$ phase diagram[30], whereas such states are absent in RMG systems with 6-8 layers[26-30, 34, 38], which may indicate a different origin of superconducting pairing from those in the semimetallic region. Several theoretical proposals have been put forward to account for superconductivity in RMG systems, for example, including purely electronic mechanisms such as isospin fluctuations near symmetry-breaking transitions [55,56] and electron-electron interaction-driven pairing via the Kohn–Luttinger mechanism, potentially enhanced by quantum geometric effects [57]. Beyond purely electronic mechanisms, electron-phonon coupling has also been proposed as an alternative origin for the superconducting phases observed in RMG [58-59]. However, a unified consensus on the microscopic mechanism of superconductivity in RMG systems has yet to be reached and remains an open question requiring further investigations. Our findings contribute to the understanding of superconductivity in RMG systems, and establish them as versatile and highly tunable platforms for probing its underlying mechanisms.

**Acknowledgement-**This work is supported by the National Key R&D Program of China (Nos. 2022YFA1402404, 2022YFA1405400, 2022YFA1402702, 2024YFA1410100, 2025YFE0201200, 2021YFA1202902), the National Natural Science Foundation of China (Nos. 12374045, 12350403, 92565302, 12488101, 12350005, 12550403, 92365302, 22325203, 12474156, 12474121, 92577203, 12574146, 12574174), the Quantum Science and Technology-National Science and Technology Major Project (Nos. 2021ZD0302500, 2021ZD0302600, 2025ZD0300500), the Science and Technology Commission of Shanghai Municipality (Grants Nos. 24QA2703700, 24LZ1401100, 2019SHZDZX01, 24LZ1401000), Cultivation Project of Shanghai Research Center for Quantum Sciences (Grant No. LZPY2024-04). C.L. is supported by T.D. Lee scholarship. X. Liu and T.L. acknowledge the Shanghai Jiao Tong University 2030 Initiative Program. X. Liu and G.C. acknowledges “Shuguang Program” supported by Shanghai Education Development Foundation and Shanghai Municipal Education Commission. T.L. acknowledges support from the Asian Young Scientist Fellowship (AYSF) and the New Cornerstone Science Foundation through the XPLORER PRIZE. G.C. acknowledges the Scientific Research Innovation Capability Support Project for Young Faculty (no. ZY2025064). K. W. and T. T. acknowledge support from the JSPS KAKENHI (Nos. 21H05233 and 23H02052) and World Premier International Research Center Initiative (WPI), MEXT, Japan.

## Figures

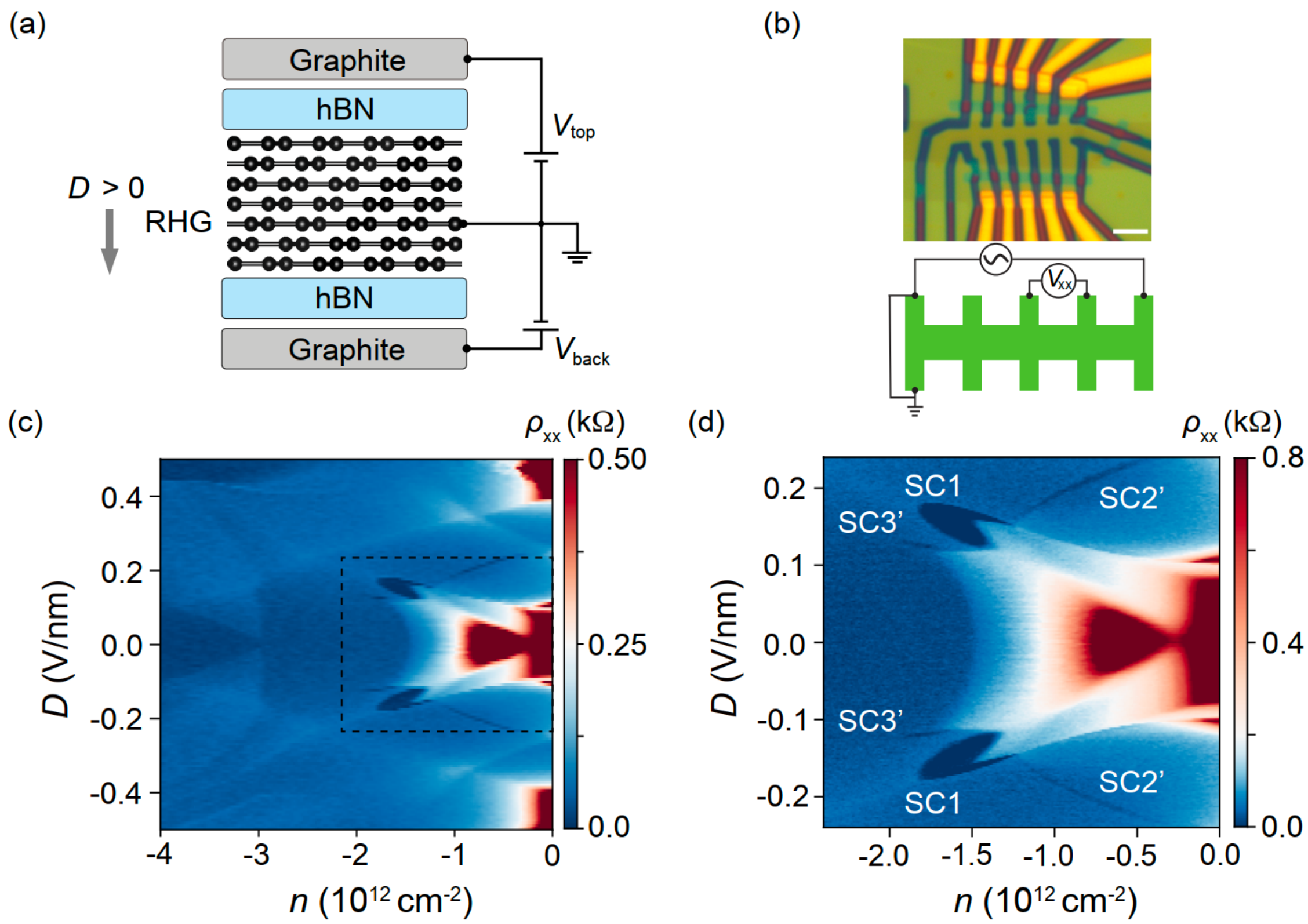

Fig.1. The transport characterization of rhombohedral heptalayer graphene (RHG). (a) Schematic of the RHG device equipped with dual graphite gates used in the measurements. (b) The optical image of the device (upper panel), with a scale bar of 3 μm. Schematic of the measurement configurations (lower panel). (c) Phase diagram of longitudinal resistivity $\rho_{xx}$ as a function of displacement electric field $D$ and charge carrier density $n$, measured at a nominal temperature $T$ = 10 mK. (d) The zoomed-in image of the region highlighted by the dashed lines in (c). One superconducting state and two signatures of superconductivity are labeled as SC1, SC2' and SC3', respectively.

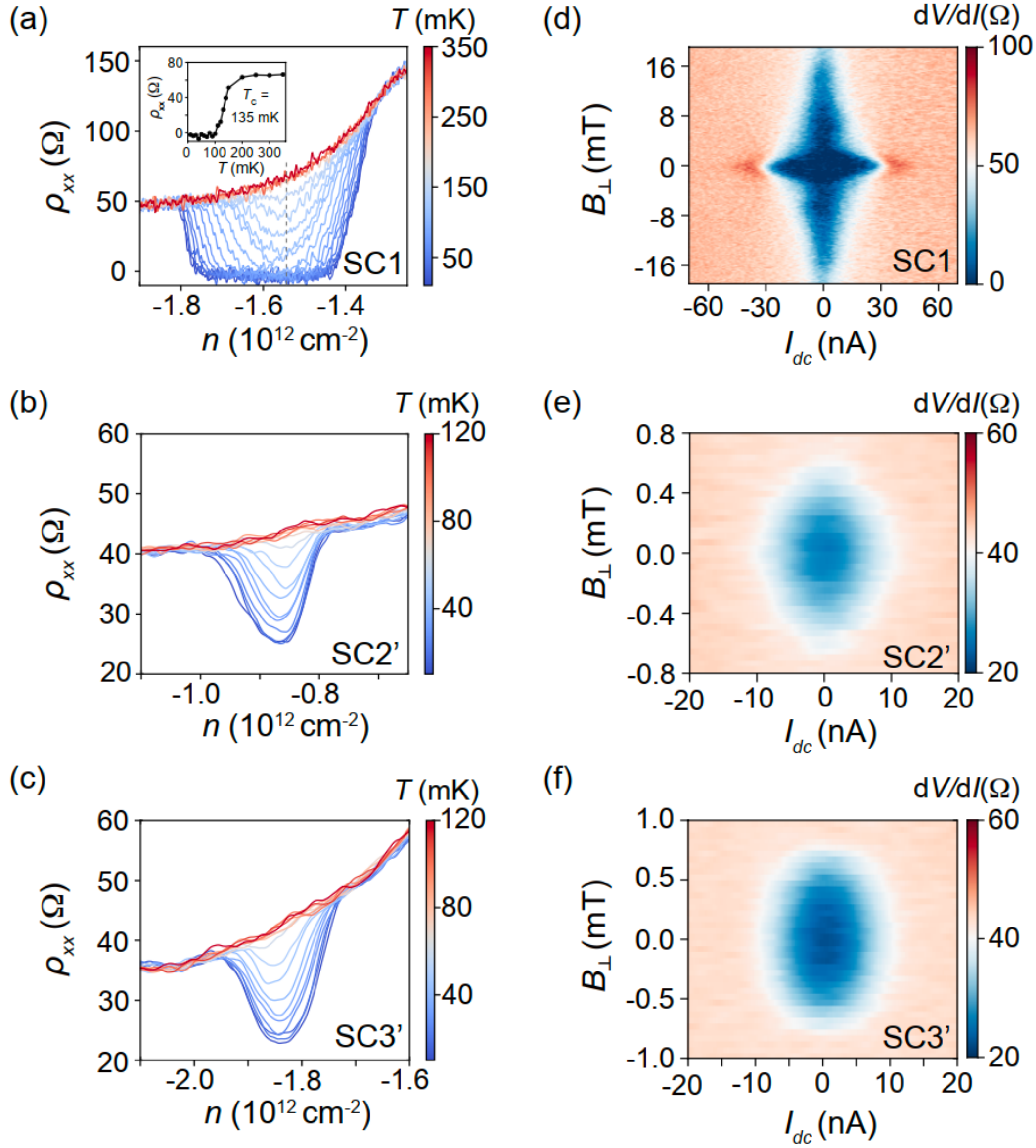

Fig.2. The transport characterization of SC1, SC2' and SC3' states in RHG. (a)-(c) The temperature dependence of $\rho_{xx}$ as a function of $n$ for SC1, SC2' (with $D$ = -0.2 V/nm) and SC3' (with $D$ = -0.12 V/nm), respectively. The inset in (a) shows $\rho_{xx}$ as a function of temperature at $n$ = -1.54 × $10^{12}$ cm$^{-2}$ and $D$ = 0.148 V/nm for SC1, and the displacement electric field $D$ in (a) is not constant, which spans from 0.12 V/nm to 0.18 V/nm; the corresponding trajectory is highlighted in Fig. S1. (d)-(f) Differential resistivity d$V$/d$I$ as a function of perpendicular magnetic field $B_\perp$ and direct current $I_{dc}$ for SC1 (with $n$ = -1.568 × $10^{12}$ cm$^{-2}$ and $D$ = 0.16 V/nm), SC2' (with $n$ = -0.86 × $10^{12}$ cm$^{-2}$ and $D$ = -0.2 V/nm) and SC3' (with $n$ = -1.84 × $10^{12}$ cm$^{-2}$ and $D$ = -0.12 V/nm), respectively.

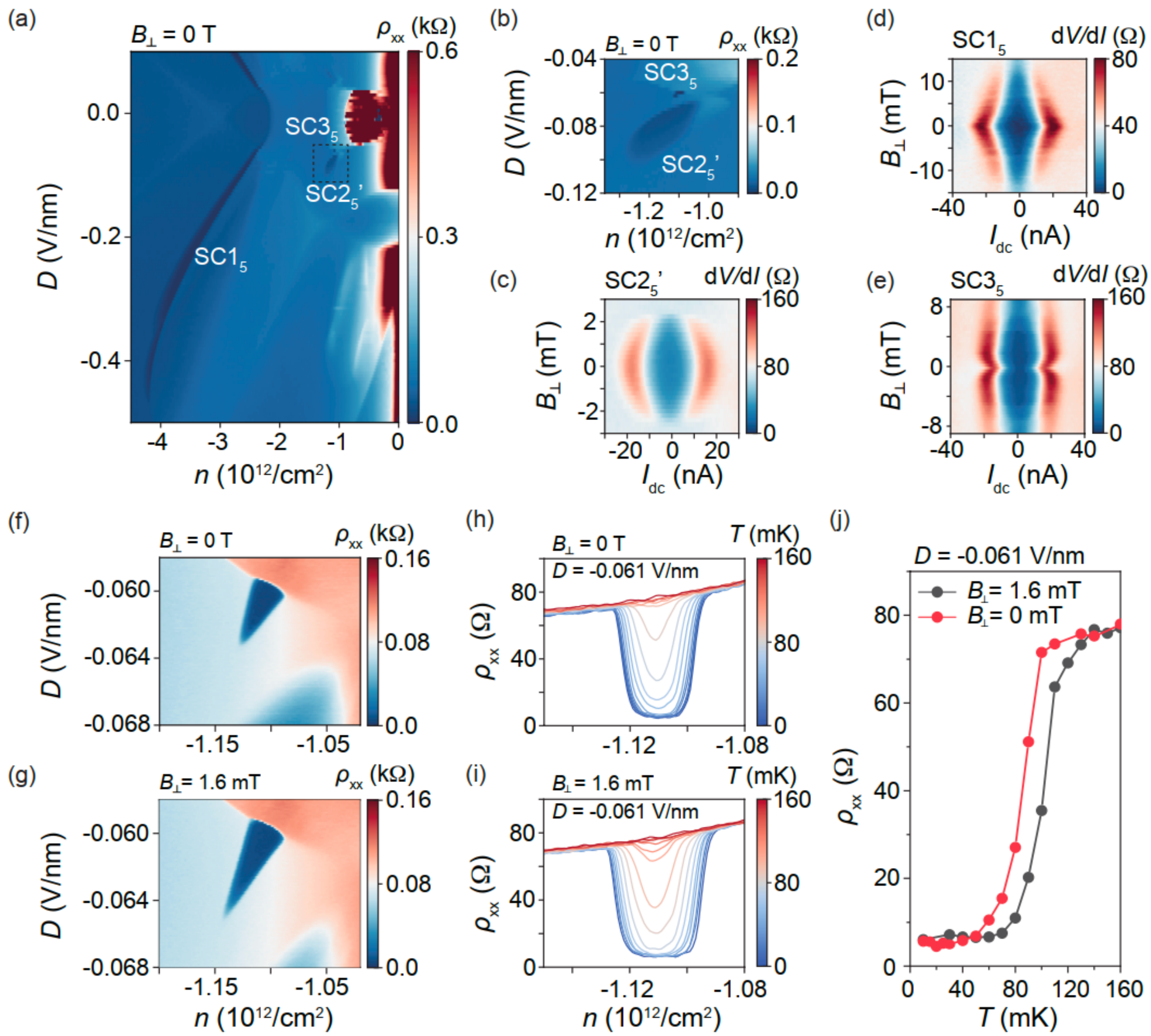

Fig. 3. Phase diagram and superconductivity in a RPG. (a) Phase diagram of $\rho_{xx}$ as a function of $n$ and $D$ at zero magnetic field. Two superconducting states, and one signature of superconductivity are observed and labeled as SC1$_5$, SC3$_5$, and SC2$_5$' respectively. (b) Zoomed-in map of the region highlighted by the dashed line in (a). (c)-(e) Differential resistivity d$V$/d$I$ as a function of $I_{dc}$ and perpendicular magnetic field $B_\perp$ for SC1$_5$, SC2$_5$' and SC3$_5$ states. (f), (g) Phase diagram of $\rho_{xx}$ as a function of $n$ and $D$ for SC3$_5$ at $B_\perp$ = 0 mT (f) and 1.6 mT (g), respectively. (h), (i) Temperature dependence of $\rho_{xx}$ as a function of $n$ at $D$ = -0.061 V/nm for SC3$_5$, measured at $B_\perp$ = 0 mT (h) and 1.6 mT (i), respectively. (j) Temperature dependence of $\rho_{xx}$ at optimal doping ($n$ = -1.11× $10^{12}$ cm$^{-2}$ for both panels (h) and (i)), extracted from panels (h) and (i) at $B_\perp$ = 0 and 1.6 mT.

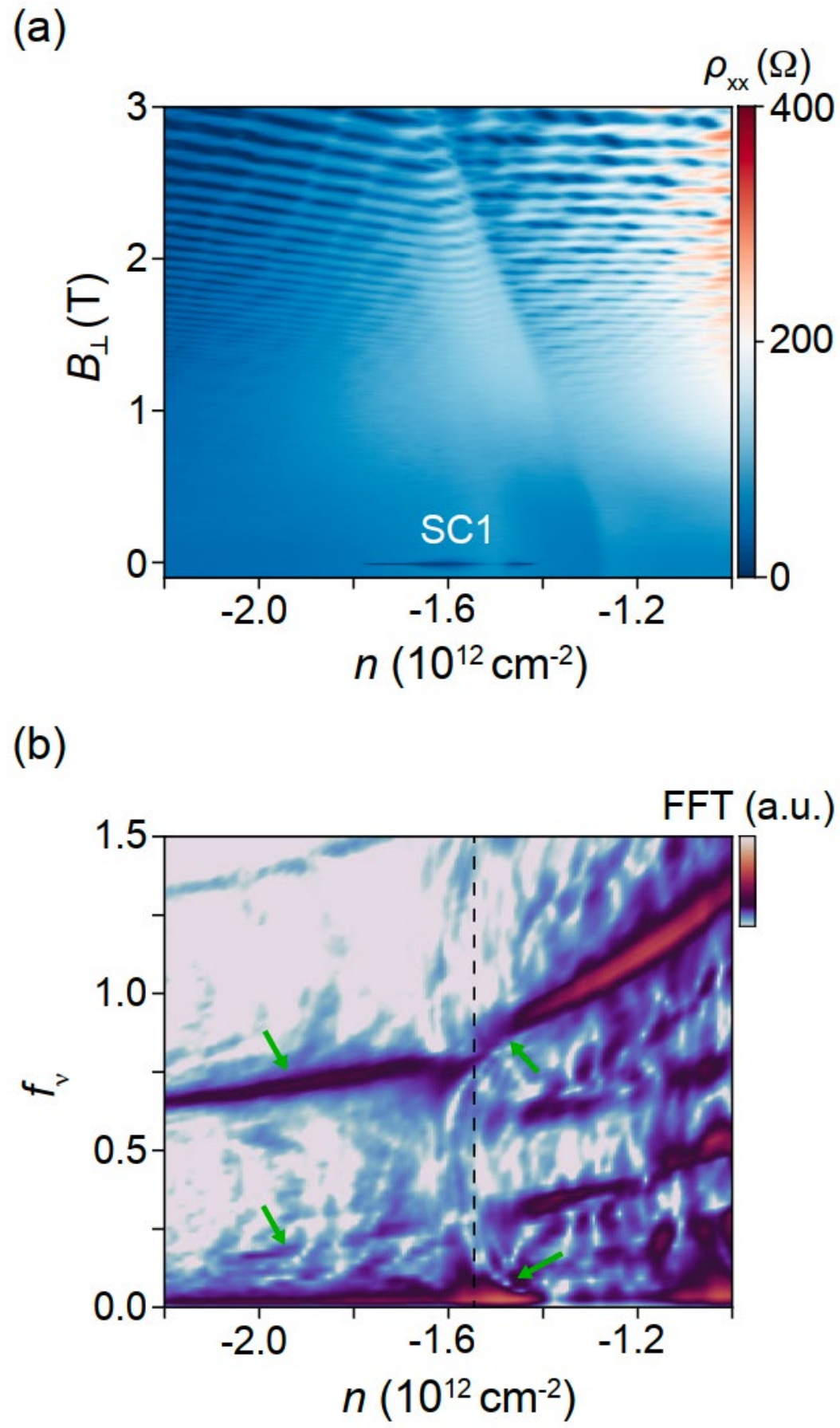

Fig. 4. Fermiology analysis for the RHG device. (a) Landau fan diagram of $\rho_{xx}$ measured at $T = 10$ mK and $D = 0.157$ V/nm. (b) The FFT analysis of the landau fan in (a), using data with 1 T < $B_\perp$ < 3 T. The prominent frequency peaks are highlighted by green arrows. The dashed line marks the boundary between two different Fermi surfaces of the normal states. Two prominent peaks in the high density region satisfy $|f_\nu^{(1)} - f_\nu^{(2)}| = 1/2$, consistent with a half metallic phase. In the lower density region, the two prominent peaks roughly satisfy $|f_\nu^{(1)} + f_\nu^{(2)}| = 1$, consistent with a $PIP_1$ phase.

# Supplementary Materials

## Device fabrication

Thin flakes of graphite and hexagonal boron nitride (hBN) were obtained by mechanical exfoliation from bulk crystals. The rhombohedral stacking order of graphene flakes was identified using infrared optical microscopy, and the rhombohedral regions were further isolated from other stacking orders by AFM lithography. Van der Waals heterostructures were assembled using a dry-transfer technique with a poly(bisphenol A carbonate) (PC) film supported on a polydimethylsiloxane (PDMS) stamp. During assembly, a top hBN flake was first picked up and subsequently used to pick up the rhombohedral graphene layers. The resulting stack was then transferred onto a pre-patterned bottom stacks consisting of a bottom hBN dielectric and a graphite gate on a Si/SiO2 substrate. After assembly, infrared optical microscopy was again employed to identify regions where the rhombohedral graphene remained structurally unrelaxed. These regions were cleaned using AFM contact-mode, after which the top gate was released onto the target area. The completed heterostructures were patterned into standard Hall bar geometries using electron-beam lithography followed by reactive ion etching with CHF3/O2. Electrical edge contacts were fabricated by electron-beam evaporation of Cr/Au (2/80 nm).

Table I Summary of superconducting properties in RHG and RPG

| States | $n$-$D$ range at $B_\perp = 0$ T ($\times 10^{12}$ cm$^{-2}$, V/nm) | $T_c$ at $B_\perp = 0$ T (mK) | $B_{c\perp}$ (mT) | Enhanced by a small $B_\perp$ | Residual resistivity at base temperature (Ω) |
|---|---|---|---|---|---|
| SC1 (RHG) | $-1.85 < n < -1.25$ $0.12 < D < 0.18$; $-1.85 < n < -1.25$ $-0.18 < D < -0.12$ | 134 (Fig.2(a)) | 12.5 (Fig.S2(a)) | No | Zero within the noise floor |
| SC2' (RHG) | $-1.2 < n < -0.3$ $0.165 < D < 0.24$; $-1.2 < n < -0.3$ $-0.24 < D < -0.162$ | — | — | No | 25 |
| SC3' (RHG) | $-1.95 < n < -1.65$ $0.117 < D < 0.131$; $-2.0 < n < -1.65$ $-0.127 < D < -0.115$ | — | — | No | 23 |
| $SC1_5$ (RPG) | $-4.3 < n < -2.3$ $-0.46 < D < -0.04$ | 91 (Fig.S4(a)) | 10 (Fig.S2(d)) | No | Zero within the noise floor |
| $SC2_5$' (RPG) | $-1.28 < n < -1.03$ $-0.1 < D < -0.065$ | — | — | No | 28 |
| $SC3_5$ (RPG) | $-1.13 < n < -1.09$ $-0.063 < D < -0.059$ | 84 (Fig. 3(h)) | — | Yes | 4 |

Note: (i) Both devices were measured in the same dilution refrigerator using the same measurement setup. (ii) The thicknesses of the top and bottom hBN layers are approximately 20 nm and 24 nm for RHG device, and 18 nm and 22 nm for RPG device, respectively.

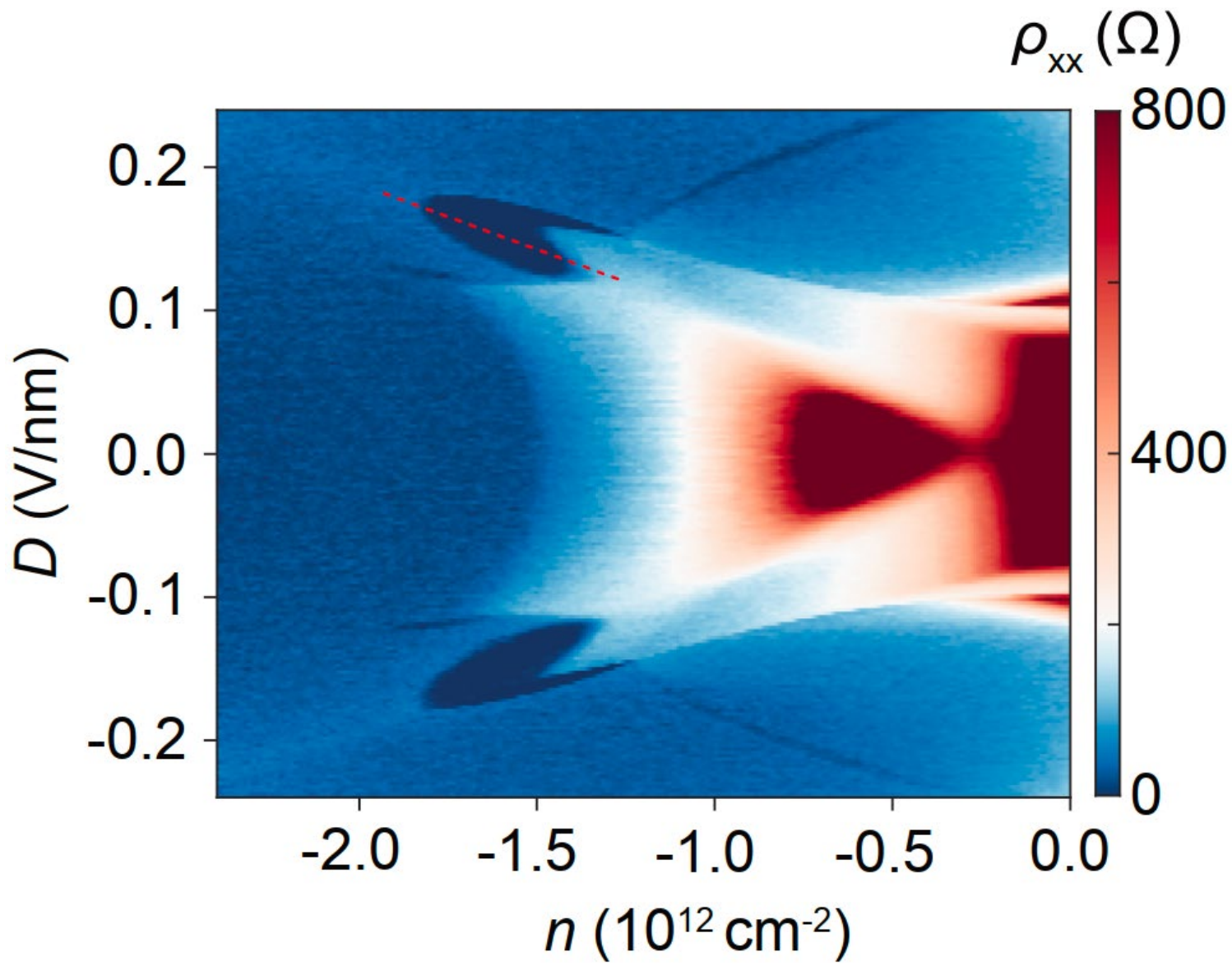

Fig. S1. The same *n*-*D* phase diagram as presented in Fig. 1(d). The red dashed line indicates the measurement trajectory shown in Fig. 2(a).

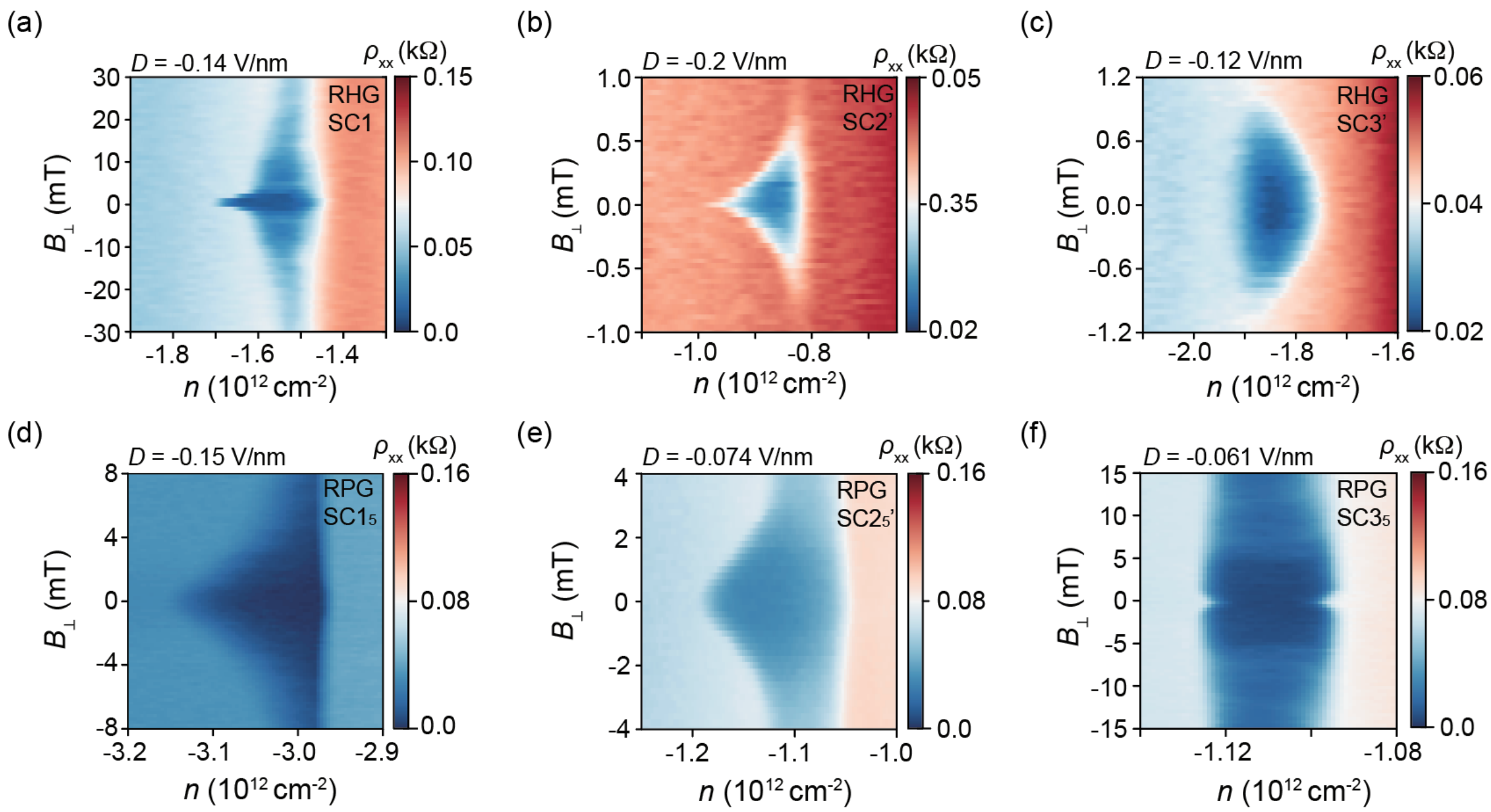

Fig. S2. The perpendicular magnetic field $B_\perp$ dependence of the superconducting states. (a) - (c) $\rho_{xx}$ maps as a function of density $n$ and $B_\perp$ in RHG for SC1 (a), SC2'(b), and SC3' (c). (d) - (f) $\rho_{xx}$ maps as a function of density $n$ and $B_\perp$ in RPG for $SC1_5$ (d), $SC2_5$' (e) , and $SC3_5$ (f) states, respectively. A non-monotonic evolution with $B_\perp$ is observed only for the $SC3_5$ state.

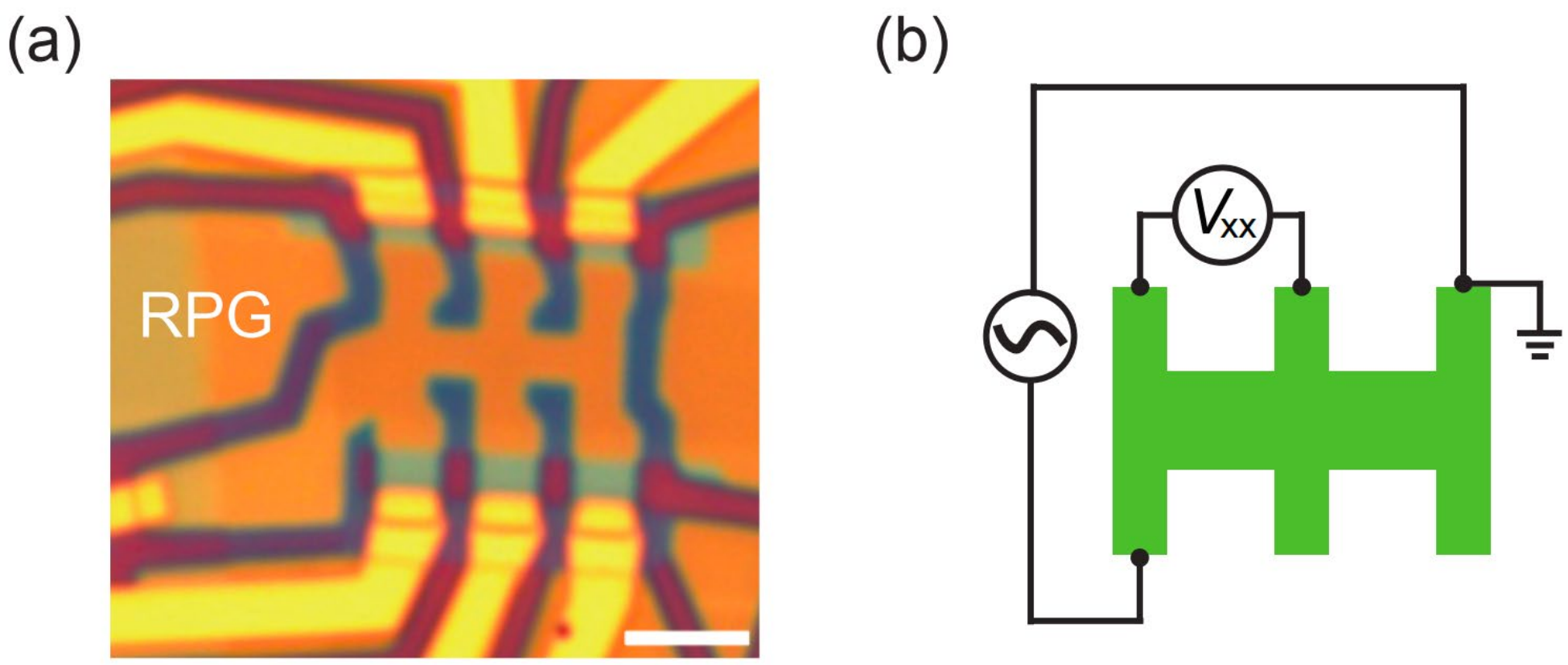

Fig. S3. Characterization and measurement setup of the RPG device. (a) Optical image of the RPG device fabricated in a Hall bar geometry. The scale bar represents 3 μm. (b) Schematic of the measurement configuration for the RPG device. Note that the schematic in (b) is oriented identically to the optical image in (a).

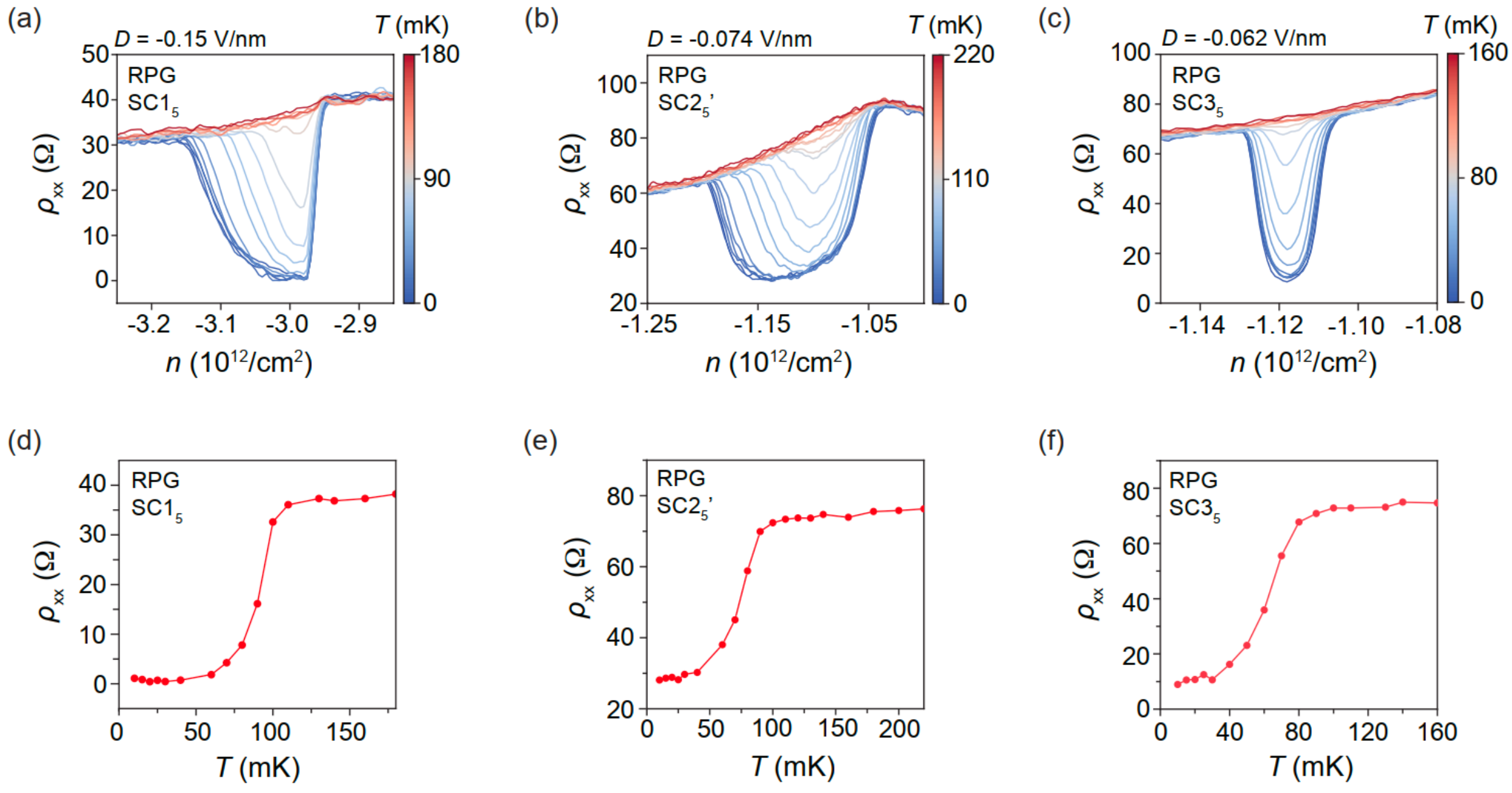

Fig. S4. Transport characterizations for $SC1_5$, $SC2_5$'and $SC3_5$ in RPG. (a)-(c) Temperature dependence of $\rho_{xx}$ as a function of density for $SC1_5$ (a), $SC2_5$' (b), and $SC3_5$ (c). (d)-(f) $\rho_{xx}$ as a function of temperature at the optimal doping for the traces in (a) - (c), respectively. The critical temperatures $T_c$ are approximately 91 mK and 60 mK for $SC1_5$ and $SC3_5$, respectively, where $T_c$ is defined as the temperature where the resistivity is 50% of the normal-state resistivity.

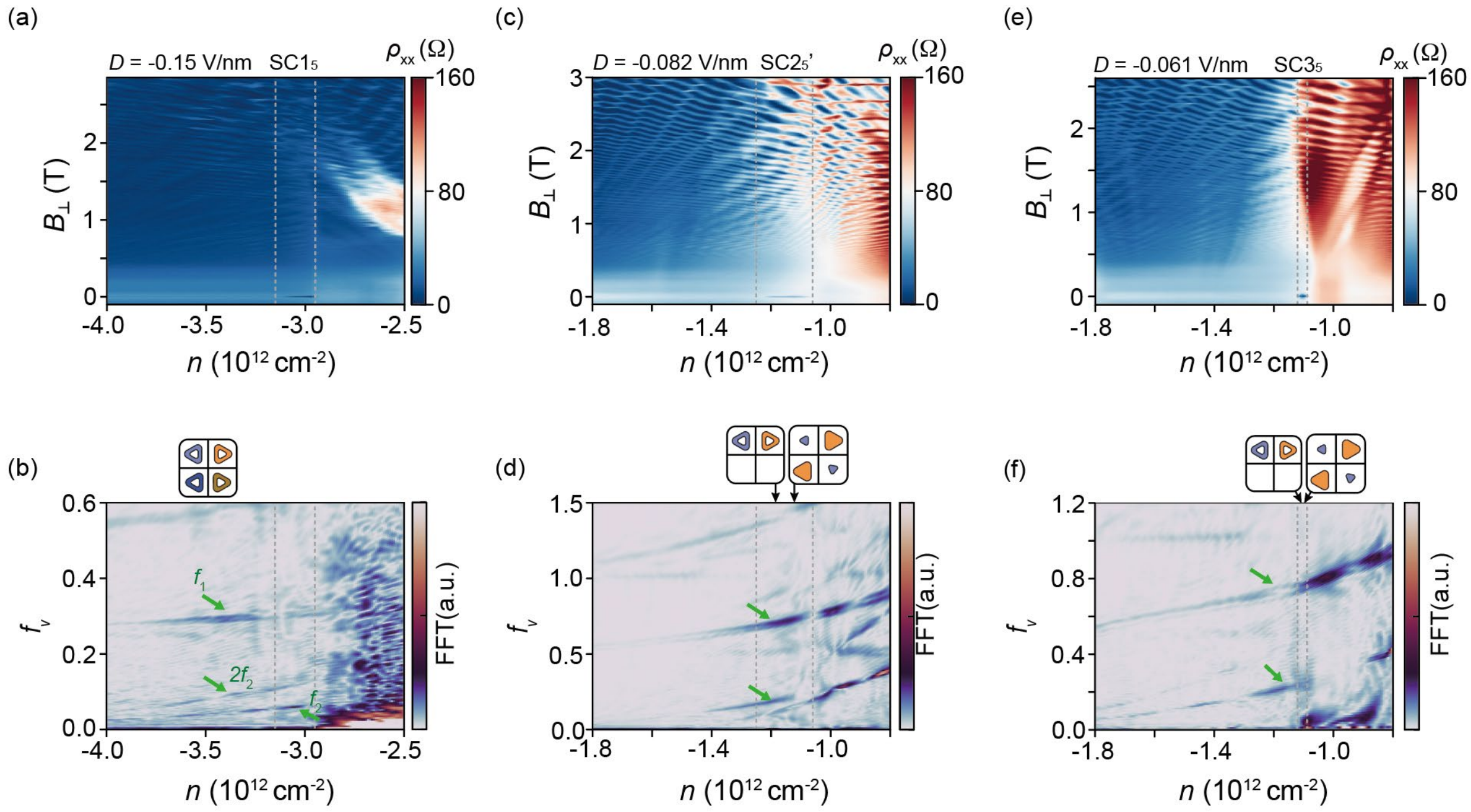

Fig. S5. Fermiology analysis of the RPG device. (a), (c), (e), Landau fan diagrams of $\rho_{xx}$ measured at $T = 10$ mK for different fixed values of $D$, as labeled in each panel accordingly. The dashed line in each panel marks the density range of $SC1_5$, $SC2_5$' and $SC3_5$, respectively. The prominent frequency peaks are highlighted by green arrows. In panel (b), the prominent frequency peak satisfies $|f_v^{(1)} - f_v^{(2)}| = 1/4$, which is consistent with a four-fold degenerate annular Fermi surface. In panels (d) and (f), $|f_v^{(1)} - f_v^{(2)}| = 1/2$, indicating that the normal state corresponds to a half-metallic phase with twofold degeneracy. The possible Fermi surface structures include a twofold-degenerate annular Fermi surface, or a configuration consisting of two electron pockets and two hole pockets.